\documentclass[11pt]{article}
\usepackage{amsmath}
\usepackage{graphicx}

\begin{document}

\title{\textbf{Spontaneous symmetry breaking by double 
lithium adsorption in polyacenes}}

\author{ Yenni. P. Ortiz$^{a}$ and Thomas H. Seligman$^{a,b}$
 \\ \small{\textit{$^a$Instituto de Ciencias F\'isicas, Universidad Nacional 
Aut\'onoma de M\'exico}},\\ \small{\textit{Cuernavaca, Morelos, M\'exico}},
\\\small{ \textit{$^b$Centro Internacional de Ciencias, Cuernavaca, Morelos, M\'exico}}}

\maketitle

\begin{abstract}
 We show that adsorption of one lithium atom to  polyacenes, i.e. 
chains of linearly fused benzene rings, will cause such chains to be slightly deformed. If we adsorb a second identical atom on the opposite side of the same ring, this deformation is dramatically enhanced despite the fact that a symmetric configuration seems possible.  
We argue, that this may be due to an instability of the Jahn-Teller type possibly indeed to a Peierls instability.
\end{abstract}

\maketitle


\section{INTRODUCTION}

Sheets, strings and tubes of hexagonal carbon structures have received an 
increasing amount of attention ever since the discovery of Fullerenes \cite{fullerenes}. In particular 
the question of adsorption of electron donors has been discussed in various contexts \cite{ant1,ant2}. The starting points of our consideration 
was a molecular study 
of small carbon sheets, treated as aromatic molecules with hydrogen 
termination at the edges \cite{js1} as well as a study on larger 
flakes \cite{js2}. We shall focus on one effect mentioned in \cite{js1}, 
namely the deformation of such structures under adsorption of one and two 
lithium atoms.  
In \cite{js2} it was pointed out, that such deformations are far 
more pronounced if we have approximately symmetric adsorption on 
both sides of the sheet. By considering a chain, i.e. a polyacene, 
we simplify the 
problem and this will allow us to get a clearer view and better understanding of the
 reasons for the symmetry breaking.

We perform numerical calculations using one standard code 
both on the Hartree Fock (HF) level and on the density functional theory (DFT) level.
 As in \cite{js1} we find qualitative agreement between both calculations, 
which indicates that the deformations observed in our calculations, 
are not too sensitive to the accuracy of the method used, yet differences 
in the deformation angle will be seen. Indeed the symmetry breaking deformation 
occurs even more clearly 
in polyacenes, and evidence points to a Jahn-Teller 
\cite{jahn-teller} like effect. 
We briefly revisit the proof of 
Ruch and Schoenhofer \cite{rs} to see, that this is plausible. As the system is 
quasi one-dimensional this effect might be reduced to a somewhat unusual 
Peierls distortion\cite{peierls}.  

The paper will be organized as 
follows: In the next section we shall discuss the configurations 
we shall use and the numerical tools that will be applied to 
obtain the results which we present in the third section. 
In the forth section we comment our findings from a group theoretical 
point of view and finally we shall give an 
outlook on the implications for other systems and on planned
additional work.

\section{CONFIGURATIONS AND COMPUTATIONAL TOOLS}

\begin{figure}[h]
\label{benVII}
 \includegraphics[height=0.3\textheight]{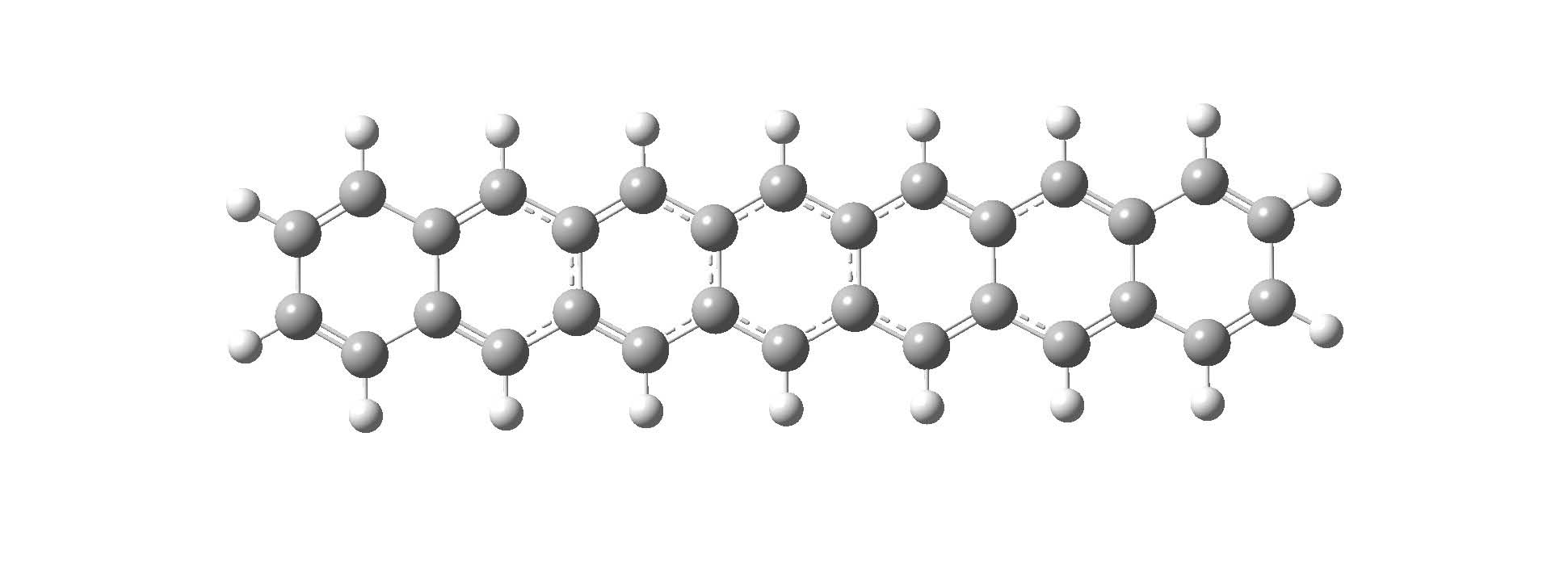}
 \caption{The heptacene molecule}
\end{figure}

As a simpler way to attack the problems of symmetry breaking discussed in 
\cite{js1,js2} we shall focus on polyacenes or acenes, i.e. chains of simple
benzene like hexagonal carbon structures bounded by hydrogen atoms. Though we calculated
chains with length from three (anthracene) to nine rings (nonacene),
we shall concentrate on showing heptacene, i.e. 
chains of seven rings as illustrated in Figure \ref{benVII}. Other polyacenes 
show analogous results using the methods described below. 
The structure is flat (not shown) and deformations of the rings at the ends
of the chain are small and imperceptible on the figure. For pentacene an atomic-force 
microscopy picture is available \cite{pentacene}.

The purpose of the calculation will be to analyze the 
stability of such configurations when lithium atoms are adsorbed.
Note that practically speaking structures beyond pentacene are very reactive, 
but this does not concern us here.
We shall keep in mind, that in  \cite{js1,js2} such deformations 
where greatly enhanced by adsorbing a lithium atom on opposite sides 
of the same carbon ring. Calculations for the larger configurations 
are quite demanding, and we have used the super-computing 
facilities of DGSCA of the University of Mexico (UNAM).

\noindent The calculations were performed with the GAUSSIAN09 program 
codes \cite{gaussian}. Geometry optimizations were performed with 
the Hartree-Fock (HF) method \cite{hartree,fock} as well as with 
DFT \cite{dft} and the agreements and differences of the
two methods will be discussed.

\noindent We have chosen the 3-21G* 
basis in which the 1s AO (Atomic Orbitals) of a first-row 
element is represented by a fixed combination of 3 GTO's 
(orbitals are approximated by a linear combination of gaussian 
basis functions), the 2s (2px etc.) are approximated by a fixed 
combination of 2 GTO's and the extra valence orbitals 2s'(2px' etc.) 
are just one GTO. The 3-21G basis set offers a reasonable compromise 
between computational burden and quality of results. Calculations 
carried beyond the Hartree Fock level by density functional theory (DFT) 
computations. The B3LYP method \cite{becke} was used both with the above basis sets and with 
the basis sets 6-31G* (6-31G + d-functions for first row atoms). This method uses 
the Becke three parameter hybrid functionals which are composed of 
$\rm A*{\rm E_{X^{Slater}}}+(1-A)*{\rm E_{X^{HF}}}+ B* \Delta {\rm E_{X^{Becke}}+E_{C^{VWN}}} +C*\Delta
{\rm E_C^{non-local}}$   
by which $\rm A, B$ and $\rm C$ are the constants determined by Becke and suitable 
for the G1 molecular set.

\section{SINGLE AND DOUBLE ADSORPTION}

We shall  now proceed to study the effects of single and double adsorption of lithium 
at a single carbon ring of heptacene. Double adsorption to the same ring is known to be 
possible only if the two adsorbed atoms are at opposite sides. We shall mainly show 
density functional calculation with the 6-31G* basis set and we shall see that the results are qualitatively 
equal to those of Hartree Fock calculations. Note that they are also qualitatively equal and quantitatively very 
 to DFT calculations with the 3-21G* basis sets.  

\begin{figure}[h]
$\begin{array}{cc}
\includegraphics[height=0.27\textheight]{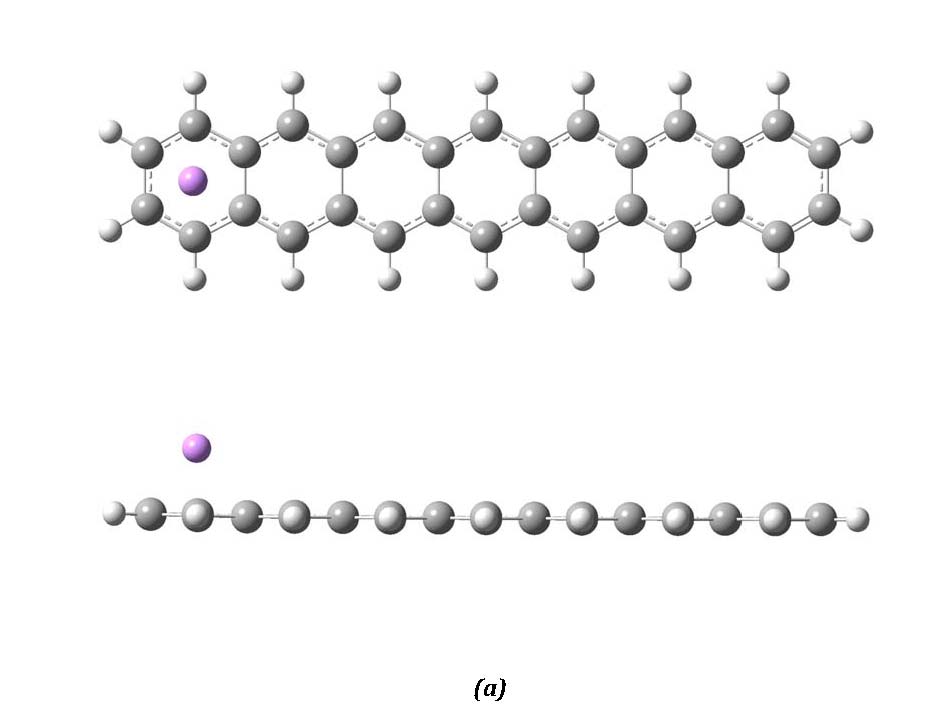}&
\includegraphics[height=0.27\textheight]{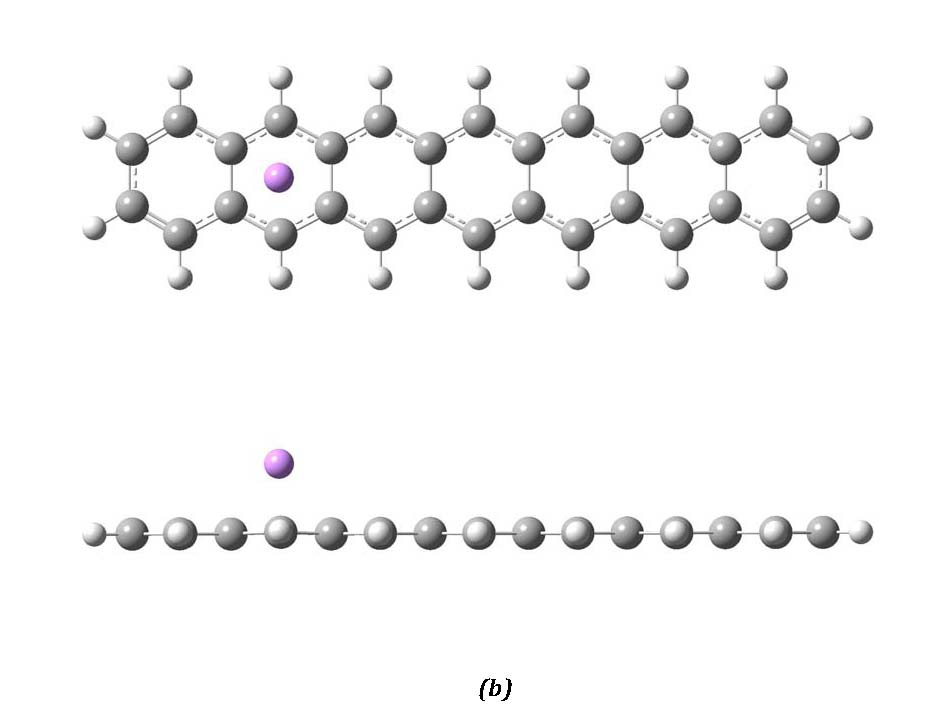}\\
 \includegraphics[height=0.27\textheight]{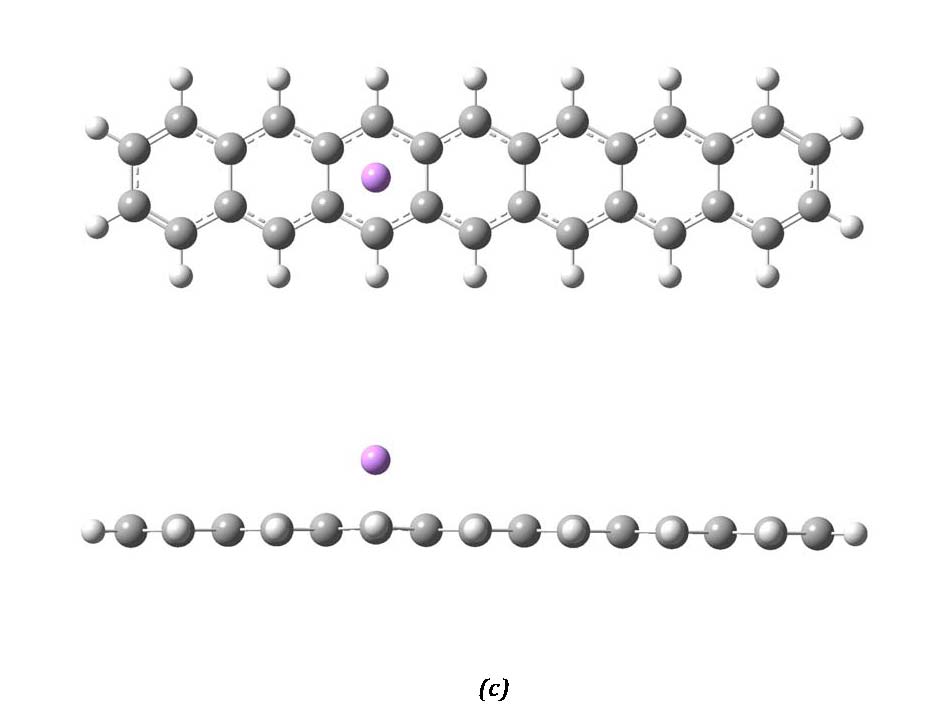}&
   \includegraphics[height=0.27\textheight]{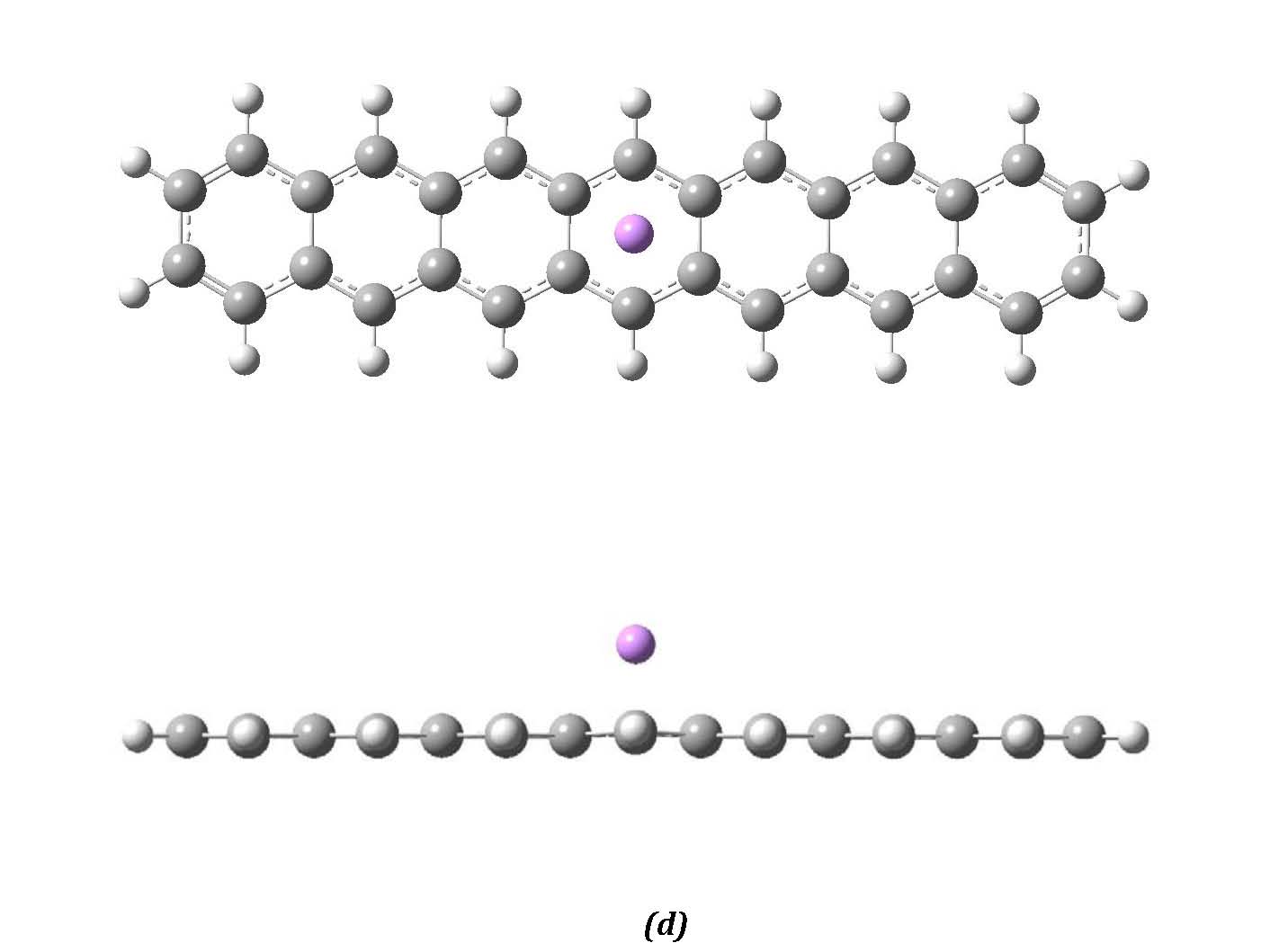}
\end{array}$
 \caption{Chain of seven rings with one Lithium atom ($a$) 
 adsorbed to the first ring,  ($b$) the second ring,  ($c$) 
the third ring and  ($d$) the fourth ring,}
\label{single}
\end{figure}

As expected for single Li adsorption 
the deformations of the molecule are slight, as seen in Fig.~\ref{single} for different choice of the ring to which we adsorb.

\noindent

We thus proceed to use these structures as starting 
points for the calculation of the adsorption of the second Lithium atom to the 
other side of same the ring to which the first Lithium was adsorbed.
 
 \begin{figure}[h]
$\begin{array}{cc}

\includegraphics[height=0.27\textheight]{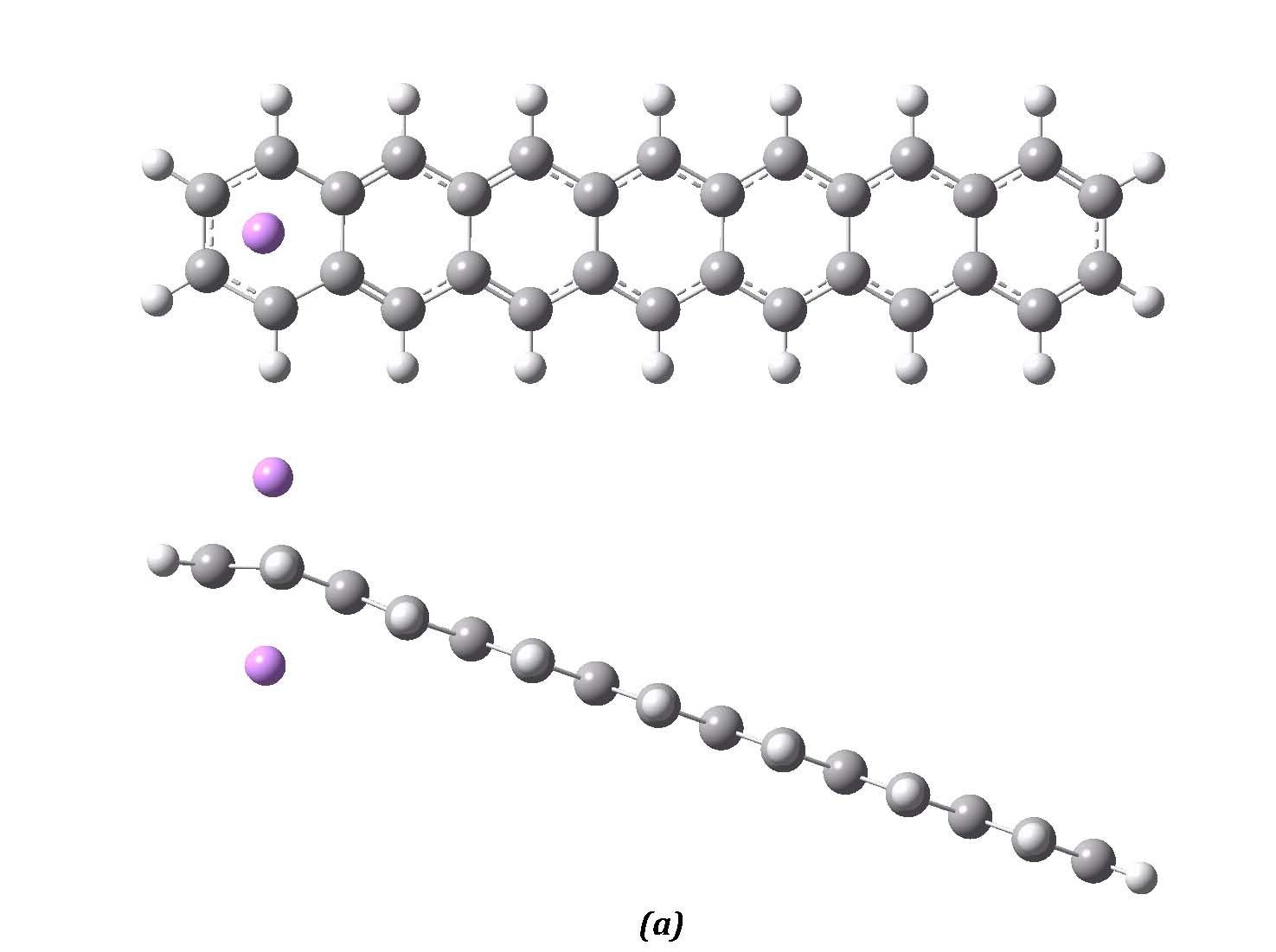}&
\includegraphics[height=0.27\textheight]{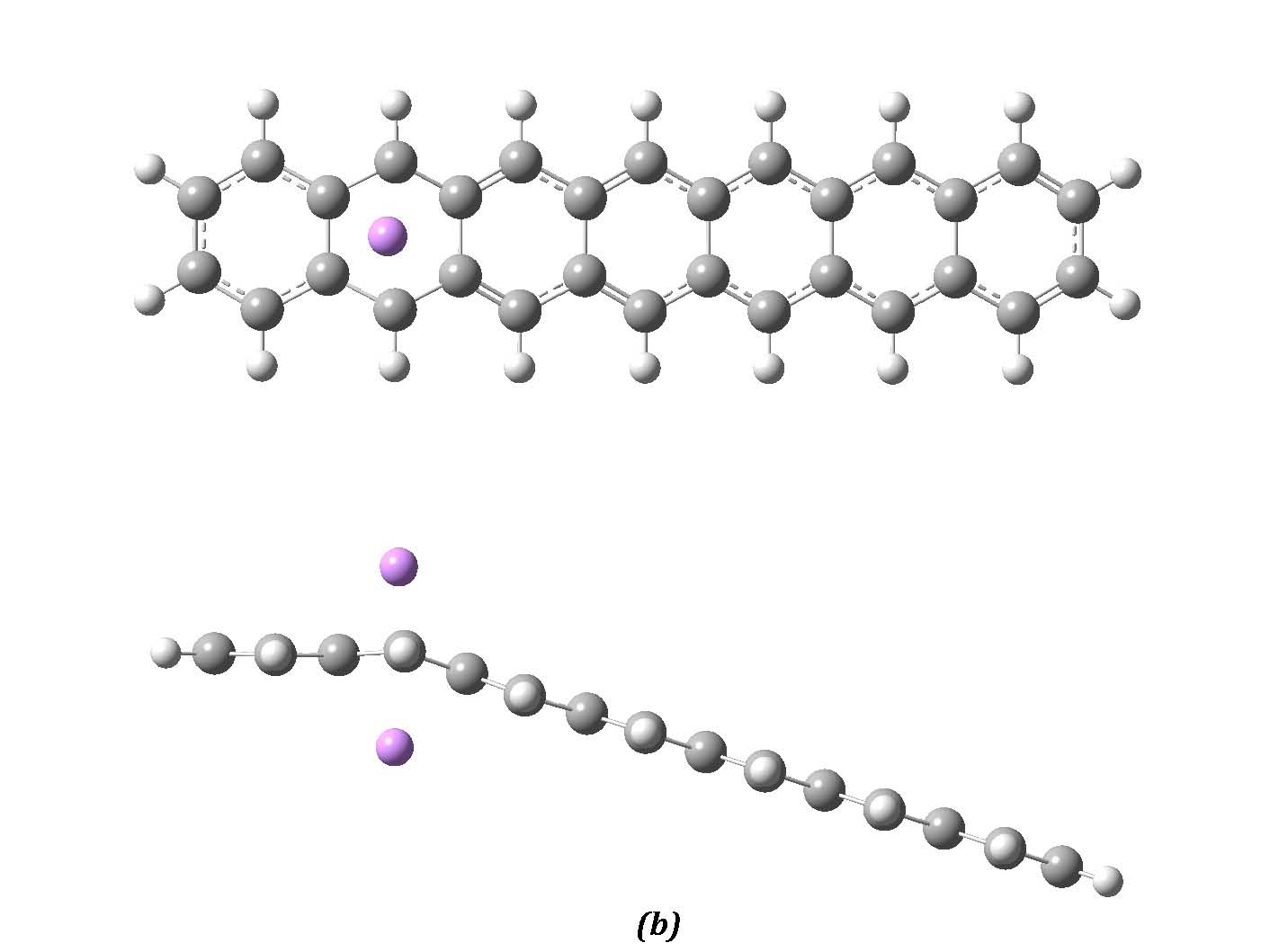}\\
 \includegraphics[height=0.27\textheight]{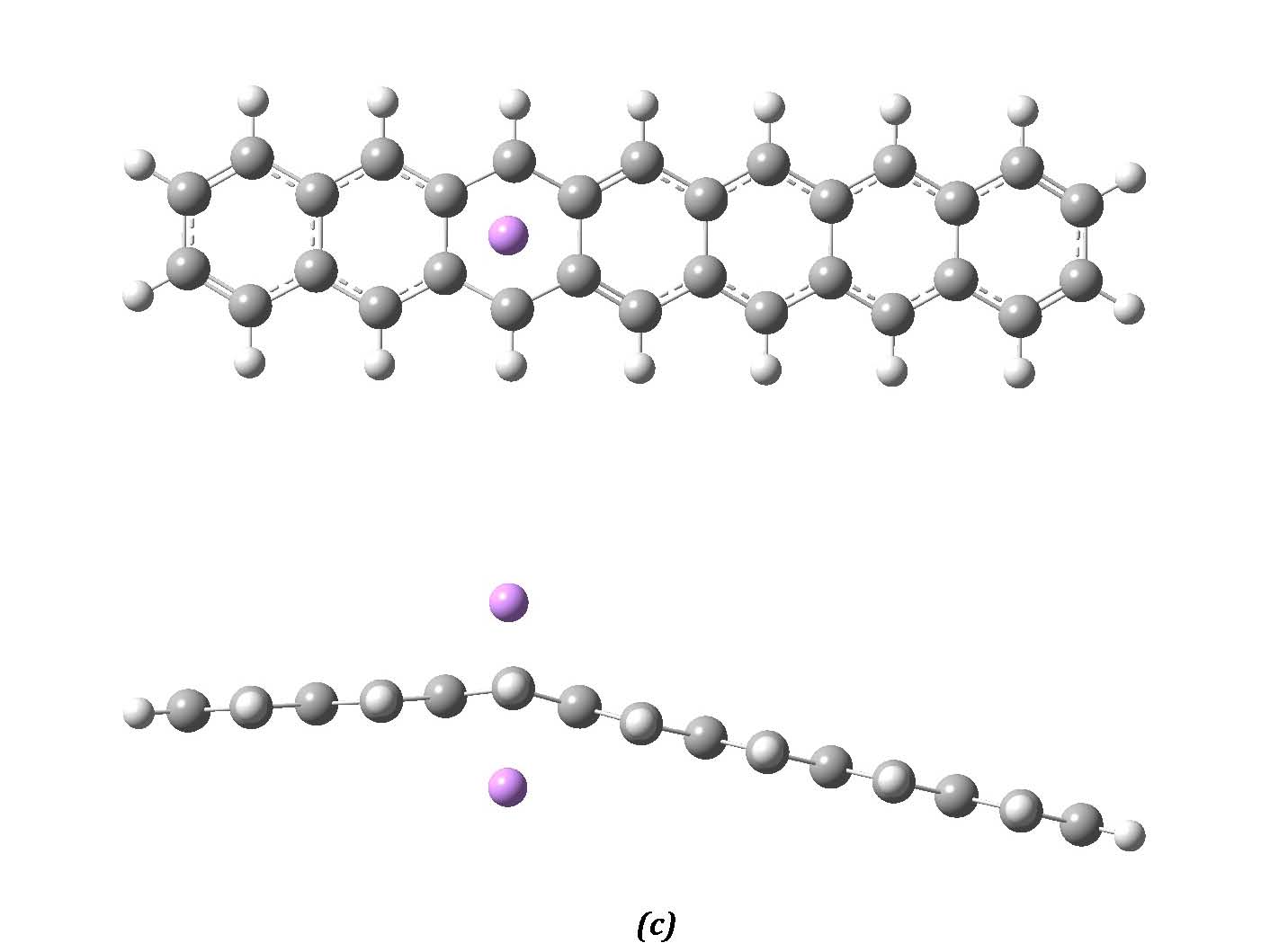}&
 \includegraphics[height=0.27\textheight]{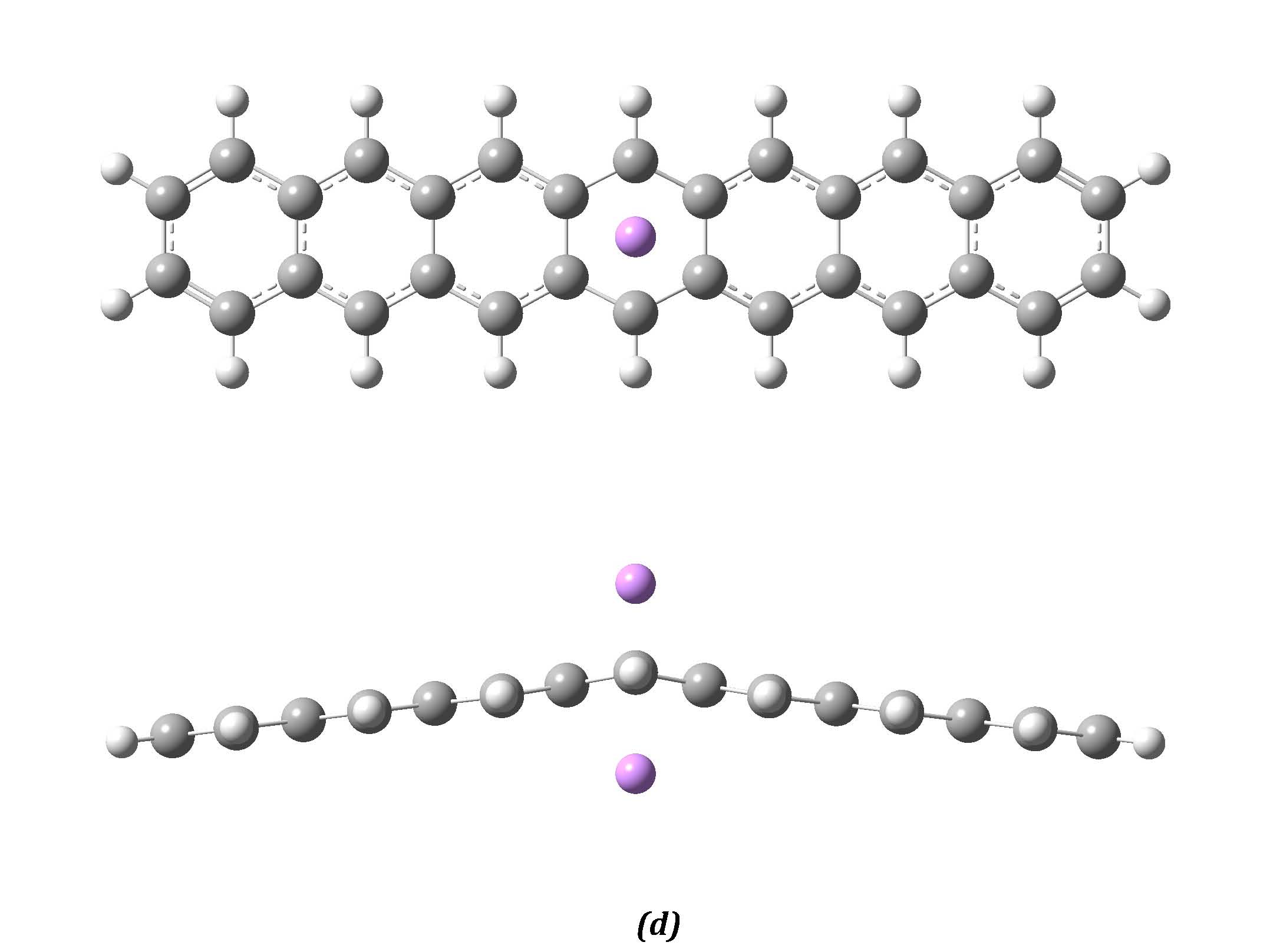}
\end{array}$
 \caption{Chain of seven rings with two Lithium atoms ($a$) 
 adsorbed to the first ring,  ($b$) the second ring,  ($c$) 
the third ring and  ($d$) the fourth ring,}
\label{surp} 
\end{figure}

\noindent The result shown in Fig. \ref{surp} is spectacular. The deformation has increased 
greatly, and as compared to \cite{js1,js2} it is much better defined, because the 
effect on a chain, i.e. on a polyacene is not local but global. If this were sheet, strong bonds of far lying rings would maintain global properties. Thus only 
a local distortion could occur.  As a symmetric structure seems possible, what we see is clearly spontaneous symmetry breaking of one kind or another. By first depositing one Lithium atom 
which produces a slight deformation we determine to which side the chain will 
bend. This was tested explicitly, but we will not show the corresponding calculation, 
as the result is rather obvious. Yet, as a caveat, we mention that, if we add 
both lithium atoms simultaneously, the "Gaussian" code does not always find the asymmetric configuration despite of its energy advantage. 
Instead it can remain in a symmetric configuration. This seems to point to a small local minimum of energy. This will be explored in a future publication.

\begin{figure}[h]
$\begin{array}{cc}
\includegraphics[height=0.27\textheight]{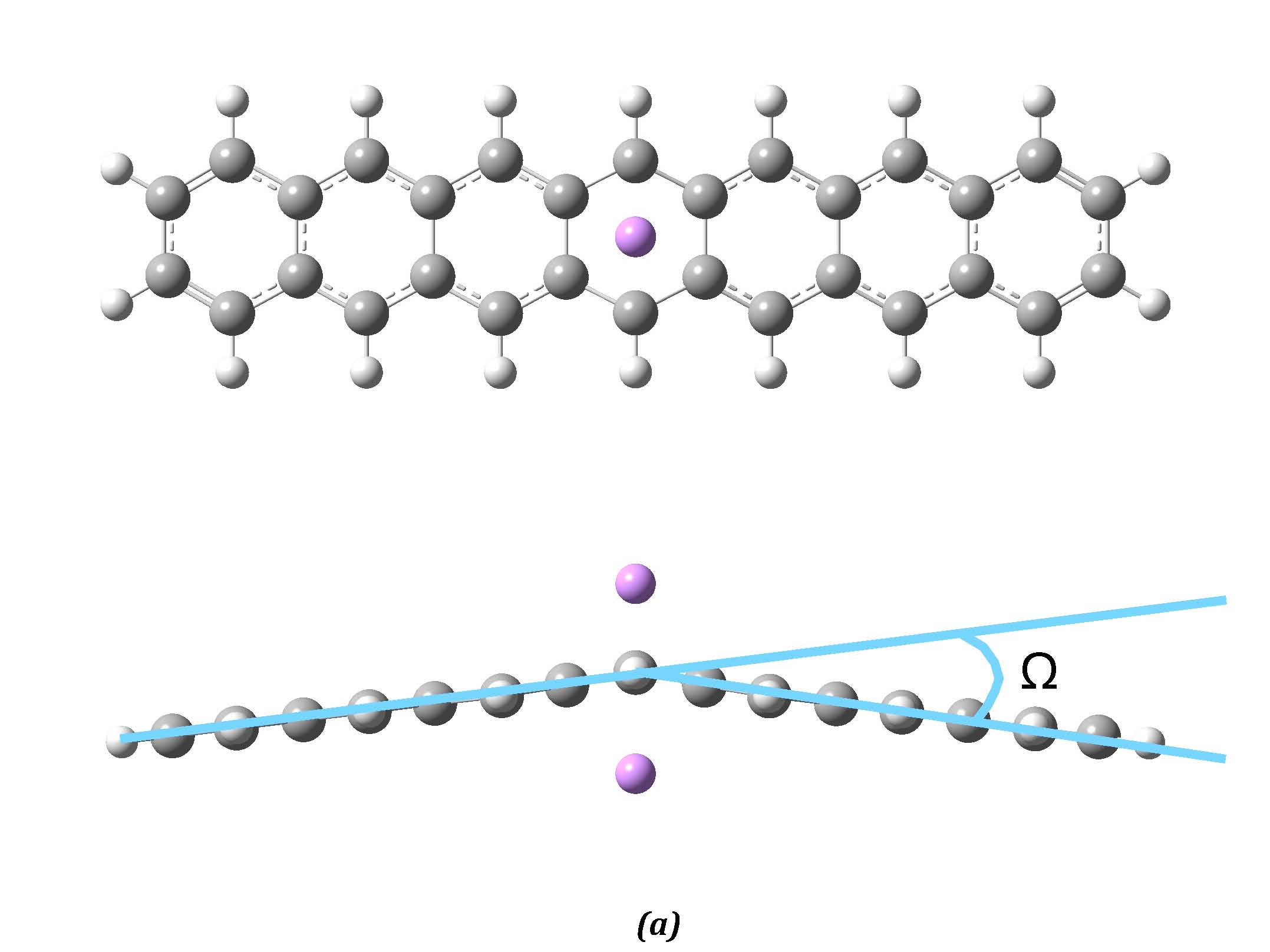}&
\includegraphics[height=0.27\textheight]{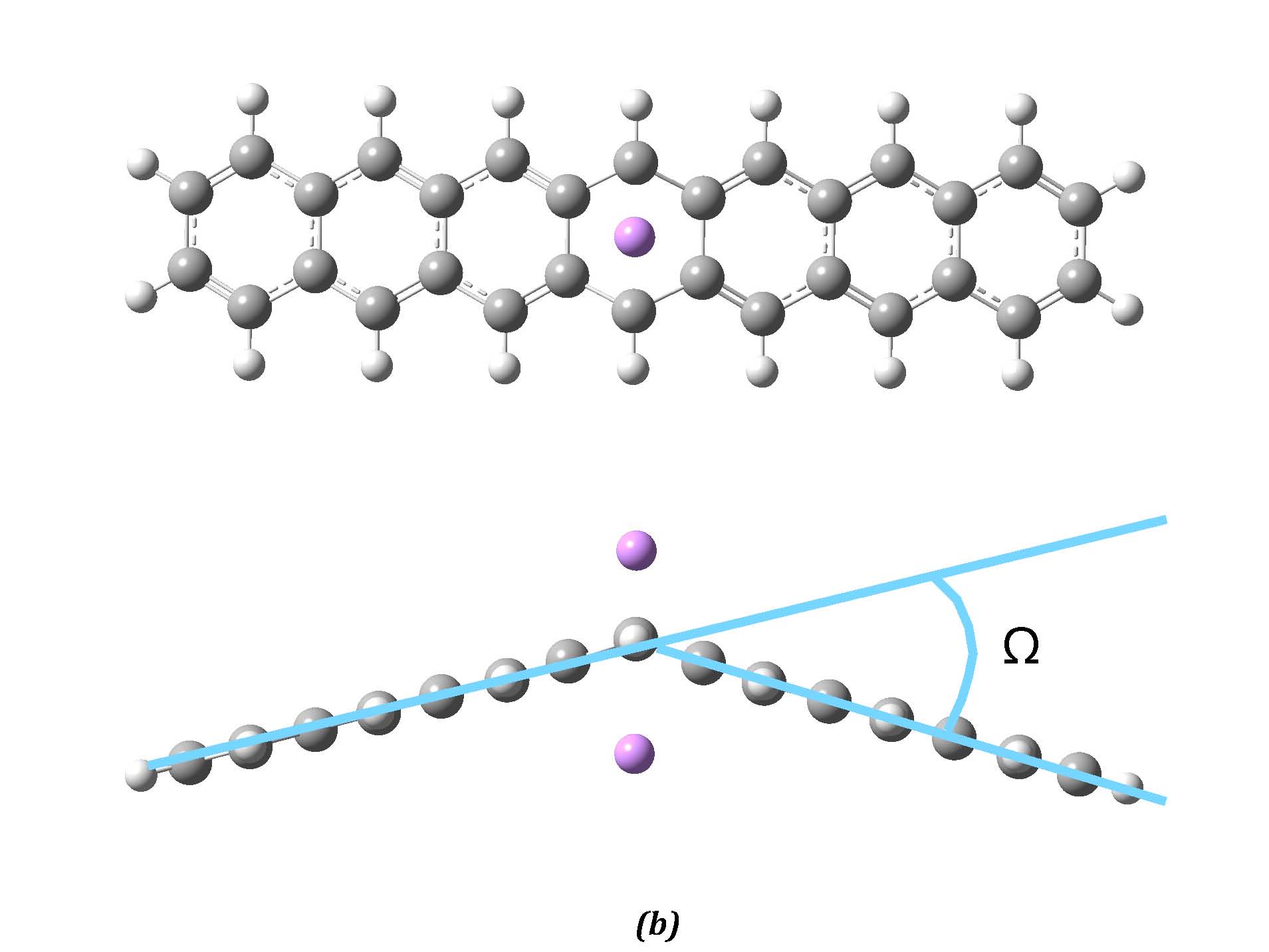}\\
\end{array}$
\caption{Heptacene with two lithium atoms adsorbed on opposite sides of the 
central ring.  ($a$) 
shows the DFT calculation and illustrates the bending angle 
 shown in table  \ref{table}. ($b$) shows the HF calculation and also illustrates the bending angle 
 shown in table  \ref{table}.}
\label{comparison}

\end{figure}

In order to demonstrate the robustness of the result we shall show in Fig. \ref{comparison} that the double adsorption case behaves qualitatively similar whether we use DFT or HF. 
In Fig \ref{comparison}$a$ we see the DFT result shown already in the previous 
figure, while in part $b$ we see the equivalent HF calculation. Clearly the angle 
between the flat configuration 
and the calculated one is bigger in the Hartree Fock case; this fact seems to indicate that 
the symmetry breaking is a mean field effect, which is actually somewhat mitigated 
by correlations. 

\noindent In table \ref{table} we show the the angle of deviation form the flat configuration 
for adsorption on rings 1,2,3 and 4 (center) both for the DFT and the HF calculations. 
This angle is illustrated for the case of position 4 in Fig~\ref{comparison}.
We notice, that the larger angles for HF as compared to DFT are a systematic feature. For HF the  angle 
diminished as we use a less symmetric configuration, while it changes little for DFT.
We made the same DFT calculations using the basis sets 3-21G* and we did not find big differences in the angles. \\

\begin{table}
\centering
\begin{tabular}{ccc}
\hline
&\textbf{Angle of deformation $\Omega$}& \\
\hline
 \textbf{POSITION} & \textbf{HF}
 & \textbf{DFT}\\
 \hline
 1&19.17&19.77\\
2 & 27.75 & 21.18 \\
3 & 33.99 & 20.33 \\ 
4 &36.29 & 18.86\\
\hline
\end{tabular}
\caption{Angles deformation from flat heptacene vs position of the 
pair of lithium adsorbed to the the chain for positions 1, 2, 3 and 4 (center).}
\label{table}
\end{table}

\begin{figure}[h]
\label{VI_dft_s3}
 \includegraphics[height=0.18\textheight]{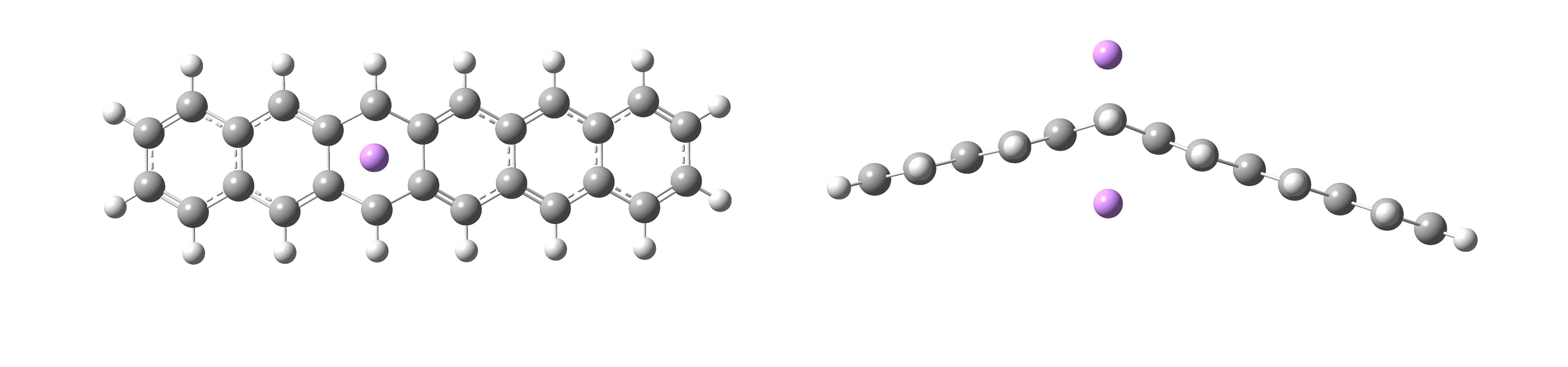}
 \caption{The hexacene molecule with two lithium atoms  adsorbed on opposite sides of the 
third ring. }
\end{figure}

If we have an even number of rings, there is no "center ring" and thus the most symmetric 
configuration does not exist.  To demonstrate that this is not relevant we show in Fig.~\ref{VI_dft_s3} one DFT (6-31G* basis sets) calculation for hexacene and see that there is no qualitative difference.  

Finally we note that adsorption of two lithium atoms to different rings,
whether they are on one side or on opposite sides,  does not display very different 
effects as compared to the adsorption of a single atom, except that the 
effect occurs on each of the sites. A detailed analysis of the stability 
of double adsorption to neighboring sites is still missing.

In view of the fact that this volume concentrates on symmetry we shall 
abstain from discussing more chemical aspects such as separation energies 
or the exact geometry of the adsorption well. As for the interesting 
question of mobility of the adsorbed atoms \cite{zhiji} we only wish to point to 
the global deformation, which would make a large scale movement necessary, if we 
move one lithium atom. 
Thus it might well be, that the movement of a correlated pair of two 
lithium atoms is more mobile then a single one. At this point, 
this is speculation, though it may be worthwhile investigating 
the question in the future. We therefore in the next section will 
try to interpret our qualitative findings by group theoretical means.

\section{A GENERALIZED JAHN-TELLER EFFECT OR PEIERLS DISTORTION? }

We briefly recall, that Jahn and Teller predict instability of the 
symmetric configuration for certain molecules or complexes  with respect 
to some point groups which might have been expected to be a symmetry groups. 
As we are dealing with a quasi 1-D system, it is worthwhile to mention, 
that the Peierls distortion is often viewed as the equivalent to the 
Jahn-Teller effect in one dimension. Yet as far a brief review of
the literature tells us, Peierls distortion  has ben considered for one-dimensional
distortions of the one-dimensional order, such as dimerization. 
Here we find for the quasi 1-D situation a deformation which bends the system, 
i.e. which deforms it perpendicular to its extension. 

We shall thus attempt 
to understand this deformation in terms of a generic derivation of the 
Jahn-Teller effect 
which was also invoked for the 2-D system \cite{js2}. Ruch and Schoenhofer 
\cite{rs} showed in a very general framework that whenever a molecular system, 
susceptible to be treated in Born-Oppenheimer approximation, had a 
symmetry group made up of a semi-direct product of two groups, we would 
find that the nuclear skeleton would be unstable and destroy the symmetry 
in one factor if the electron density would violate the symmetry of the 
other factor.  
The generalization of this 
argument is a challenge that has come up in this context, and we shall take up 
this matter in a future paper. Here we shall limit ourselves to a plausibility 
argument based on the work of Ruch and Schoenhofer, which indicates how a 
complicated, and in Moshinsky's terminology "messy" problem might be solved 
with deep and elegant group theoretical techniques.

In the present case we assume that our chains mimic 
symmetry along the chain, i.e. finite translations and reflections, 
as well as rotations by 180 degrees around the center of the chain and reflections 
in the plane of the carbon rings and perpendicular to the chain with 
the axis along the chain. The total symmetry group is a direct product 
of the 1-D symmetries in the axis of the chain and those perpendicular 
to the chain. If we take the point of view, that the main function 
of the adsorbed atoms is to donate localized electrons to the chain. 
then the electron density is no longer translationally invariant. 
This, according to the arguments of Ruch and Schoenhofer,
would lead directly to the symmetry 
breaking of the other factor in the product. Indeed this is what we observe.
The fact, that deflections are larger for centered adsorption pairs might serve a 
as a weak additional hint, as then the ends of the chain are further away and 
 the system is more symmetric to start with.

At this point we have to make two observations:

\noindent First, we deal with a direct, rather than a semi-direct product. The argument 
of \cite{rs} is rather subtle, and though it seems to carry through, we plan to 
formalize it for the special case, and to determine if it falls into the Peierls 
category.

\noindent Second in \cite{js2} an alternative rather simple explanation was offered:
Bond lengthening is natural as additional negative charge 
is available. The lengthening is lager for two than for one adsorbed lithium as the transferred charge is larger.. 
For a carbon sheet buckling is expected due to the stiff surroundings,  
but in the case of polyacenes such bond stretching in the rings 
could probably be supported without deformation. If the stretching where 
asymmetric we would expect to see torsion, which we do not see. This latter 
point is consistent with the configurations found, where the Lithium atoms 
are very nearly on an axis through the center of the ring, to which they 
are adsorbed. The alternate explanation of the spontaneous symmetry breaking is therefore not 
valid in the present case, and thus the argument for a Jahn-Teller or Peierls type behavior is much stronger than 
the one resulting from the deformations of graphene. 
This makes us confident, that we indeed see a phenomenon of this type.

\section{CONCLUSIONS AND OUTLOOK} 

After the surprising discovery of spontaneous symmetry breaking in 
calculations of carbon sheets as aromatic molecules \cite{js1,js2} 
it seemed necessary to get a better handle on the problem by 
studying a simpler system, quite in the spirit of Marcos Moshinsky. 
Unfortunately not quite in this spirit is the fact that it is still 
a numerical study only. The clear and simple numerical result in the case 
of polyacenes (chains of carbon rings with hydrogen closure) hints 
strongly to the existence of a simple explanation in terms of a Jahn-Teller or
Peierls distortion. 

The fact that symmetric adsorption of an alkaline atoms to a 
carbon ring chain causes a strong angle to be formed, is a clear and 
powerful signature of spontaneous symmetry breaking, and as such we expect a 
simple group-theoretical interpretation. We hinted, that along the  argument of 
Ruch and Schoenhofer this should be possible. Yet, this 
paper gives a very profound argumentation for the instability of the symmetric systems. 
In the present, much simpler, case we hope to be able 
to reduce the argument to a correspondingly simpler form. Thereby we also
expect to explore the full scope of the ideas proposed to generalize the Jahn-Teller 
effect by Ruch and Schoenhofer \cite{rs} as well as their 
reduction to Peierls distortions.


\section{ACKNOWLEDGMENTS}
  We thank Zhi Ji, and A. Jalbout for useful discussions, the DGCTIC and  the Superc\'omputo 
department of UNAM for the use of their hardware and software. 
Financial support is acknowledged under projects IN 114310 from PAPIIT, 
DGAPA UNAM and project 79613 of CONACyT, Mexico.

\bibliographystyle{aipproc}   

\begin{thebibliography}{10}

\bibitem{fullerenes}
 Kroto, H.W.; Heath, J.R.; O'Brien, S.C.; Curl, R.F.; Smalley, R.E., ``C60: Buckminsterfullerene'',
 \emph{ Nature \textbf{318, 14}}, pp 162-163 (1985).

\bibitem{ant1}
Jijun Zhao, Alper Buldum, Jie Han and Jian Ping Lu
\emph{Nanotechnology}, Volume 13, Number 2 (2002)

\bibitem{ant2}
Saffarzadeh Alireza,
\emph{Journal of Applied Physics}, Volume 107, Issue 11, pp. 114309-114309-7 (2010).

\bibitem{js1}
Jalbout, Abraham F.; Seligman, Thomas H. ``Spontaneous Symmetry Breaking and Strong Deformations in Metal Adsorbed Graphene Sheets'',
\emph{to be published}

\bibitem{js2} 
Jalbout, Abraham F.; Seligman, Thomas H.,
\emph{Journal of Computational and Theoretical Nanoscience}, Volume 6, Number 3 , pp. 541-544(4) (2009)

\bibitem{jahn-teller}
H. Jahn and E. Teller.,
\emph{Proc. Roy. Soc.}, A \textbf{161}(905): pp 220-235, (1937).

 \bibitem{rs}
E. Ruch and A. Schoenhofer, 
\emph{Theoret. Chem.} Acta 3; 291-304 (1965) 

\bibitem{peierls}
R.E. Peierls:  ``Quantum theory of Solids'' ,
\emph{Clarendon Press, Oxford}, (1955).


\bibitem{pentacene} 
Gross, L.; Mohn, F; Moll, N; Liljeroth, P; Meyer, G 
\emph{Science} 325 (5944): 1110. doi:10.1126/science.1176210. PMID 19713523 (2009).  


\bibitem{gaussian}
M. J. Frisch et. al., GAUSSIAN09,
 \emph{Revision A.02}, Gaussian Inc. (2009).
 
\bibitem{hartree} 
D. I. Hartree,
\emph{Proc. Uamb. Phil. Soc.}, 
24, 111(1928).
 
 \bibitem{fock}
 V. Fock, 
 \emph{Zeits. f. Physik.},  61, 126 (1930).
 
 \bibitem{dft}
 Hohenberg, Pierre; Walter Kohn,  ``Inhomogeneous electron gas''
 \emph{.Physical Review}, 136 (3B): B864ÐB871.
 
 \bibitem{becke}
 A. D. Becke, 
 \emph{J. Chem. Phys.} 98 (1993), p. 5648.
 
 \bibitem{zhiji}
 Zhi Ji, A. Jalbout,
 \emph{in preparation}
 



\end{thebibliography}

\end{document}